%% file: main.tex
\documentclass{article}

\usepackage{arxiv}

\usepackage[utf8]{inputenc} 
\usepackage[T1]{fontenc}    

\usepackage{url}            
\usepackage{booktabs}       
\usepackage{amsfonts}       
\usepackage{nicefrac}       
\usepackage{microtype}      
\usepackage{lipsum}		
\usepackage{graphicx}
\usepackage{doi}

\usepackage[T1]{fontenc}
\usepackage{graphicx}
\usepackage{amsmath}
\usepackage{amsfonts}
\usepackage{amssymb}
\usepackage[justification=centering]{caption}
\usepackage{subcaption}
\usepackage{blindtext}
\usepackage{algorithm}
\usepackage{algpseudocode}
\usepackage[dvipsnames]{xcolor}
\usepackage{comment}
\usepackage{float}
\usepackage{xcolor}
\usepackage{pgfplots}
\usepackage{xparse}
\usepackage{wrapfig}
\usepackage{todonotes}
\usepackage{multirow}
\usepackage{caption}
\usepackage{lipsum}
\usepackage{booktabs}
\usepackage{stmaryrd}
\usepackage{listings}

\newcommand{\etal}{\textit{et al.}}

\title{A Practical Introduction to Side-Channel Extraction of Deep Neural Network Parameters}

\date{}

\author{ {Raphaël Joud, Pierre-Alain Moëllic, Simon Pontié} \\
	CEA Tech, Centre CMP, Equipe Commune CEA Tech - Mines Saint-Etienne, F-13541 Gardanne, France \\
	Univ. Grenoble Alpes, CEA, Leti, F-38000 Grenoble, France\\
	\texttt{\{name\}.\{surname\}@cea.fr} \\
	\And
	{Jean-Baptiste Rigaud} \\
	Mines Saint-Etienne, CEA, Leti, Centre CMP,\\ F-13541 Gardanne France\\
	\texttt{rigaud@emse.fr} \\
}

\renewcommand{\headeright}{(arxiv version)}
\renewcommand{\undertitle}{(arxiv version)}
\renewcommand{\shorttitle}{A Practical Introduction to SCA Extraction of DNN Parameters}

\begin{document}
\maketitle

\begin{abstract}
Model extraction is a major threat for embedded deep neural network models that leverages an extended attack surface. Indeed, by physically accessing a device, an adversary may exploit side-channel leakages to extract critical information of a model (i.e., its architecture or internal parameters). Different adversarial objectives are possible including a fidelity-based scenario where the architecture and parameters are precisely extracted (\textit{model cloning}). We focus this work on software implementation of deep neural networks embedded in a high-end 32-bit microcontroller (Cortex-M7) and expose several challenges related to fidelity-based parameters extraction through side-channel analysis, from the basic multiplication operation to the feed-forward connection through the layers. To precisely extract the value of parameters represented in the single-precision floating point IEEE-754 standard, we propose an iterative process that is evaluated with both simulations and traces from a Cortex-M7 target.           
To our knowledge, this work is the first to target such an high-end 32-bit platform. Importantly, we raise and discuss the remaining challenges for the complete extraction of a deep neural network model, more particularly the critical case of biases.
\end{abstract}

\section{Introduction}
\label{introduction}
\input{src/introduction.tex}

\section{Background}
\label{Background}
\input{src/theorical_background.tex}

\section{Related Work}
\label{related_works}
\input{src/related_works}

\section{Scope and Contributions}
\label{contribution}
\input{src/contribution}

\section{Experimental setup}
\label{expsetup}
\input{src/experimental_setup}

\section{Threat model}
\label{threat_model}
\input{src/threat_model.tex}
\input{figures/traces_nn_depth}

\section{Challenges and overall methodology}
\label{challenges}
\input{src/introduction_and_problems}

\section{Extraction method and experiments}
\label{xtr_method_exp_results}
\input{src/Experiences_Results/extraction_process}
\input{src/Experiences_Results/targeting_single_neuron}

\input{src/Experiences_Results/targeting_few_neurons}

\input{src/Experiences_Results/targeting_few_layers}

\section{Future Works}
\label{future_works}
\input{src/future_works}

\section{Conclusion}
\label{mitigation}
\input{src/conclusion}

\section*{Acknowledgements}
\label{acknowledgements}
\input{src/acknowledgements}

\bibliographystyle{unsrt}
\bibliography{bib/biblio}  

\end{document}

%% file: src/introduction.tex
Deep Neural Network (DNN) models are widely used in many domains with outstanding performances in several complex tasks. Therefore, an important trend in modern Machine Learning (ML) is a large-scale deployment of models in a wide variety of hardware platforms from FPGA to 32-bit microcontroller. However, major concerns related to their security are regularly highlighted with milestones works focused on availability, integrity, confidentiality and privacy threats. Even if adversarial examples are the flagship of ML security, confidentiality and privacy threats are becoming leading topics with mainly \textit{training data leakage} and \textit{model extraction}, the latest being the core subject of this work. 

\paragraph{\textbf{Model extraction.}}
The valuable aspects of a DNN model gather  
its architecture and internal parameters finely tuned to the task it is dedicated to. These  
carefully crafted parameters represent an asset for model owners and generally must remain secret. Jagielski \etal~introduce an essential distinction between the objectives of an attacker that aims at extracting the parameters of a target model~\cite{jagielski_high_2020}, by defining a clear difference between \textbf{fidelity} and \textbf{accuracy}:
\begin{itemize}
    \item \textit{Fidelity} measures how well extracted model predictions match those from the victim model. In that context, an adversary aims to precisely extract model's characteristics in order to obtain a \textit{clone model}. In such a scenario, the extraction precision is important. Additionally to model theft, the adversary may aim to enhance his level of control over the system in order to shift from a black-box to a white-box context and design more powerful attacks against the integrity, confidentiality or availability of the model. 
    \item \textit{Accuracy} aims at performing well over the underlying learning task of the original model: the attacker's objective is to steal the \textit{performance} of the model and, effortlessly, reach equal or even superior performance. In such a case, a high degree of precision is not compulsory.
\end{itemize}

\paragraph{\textbf{Attack surface.}}

The large-scale deployment of DNN models raises many security issues. Most of the studied attacks target a model as an abstraction,  
exploiting theoretical flaws. However, implementing a model to a physically accessible device open doors toward a new attack surface taking advantage of physical threats~\cite{dumont2021overview} \cite{joud2021review}, like side-channel (SCA) or fault injection analysis (FIA). This work is focused on fidelity-oriented attack targeting model confidentiality using SCA techniques.

\paragraph{\textbf{Structure of the paper.}}

\begin{figure}[t!]
	\centering
	\includegraphics[width=0.9\textwidth]{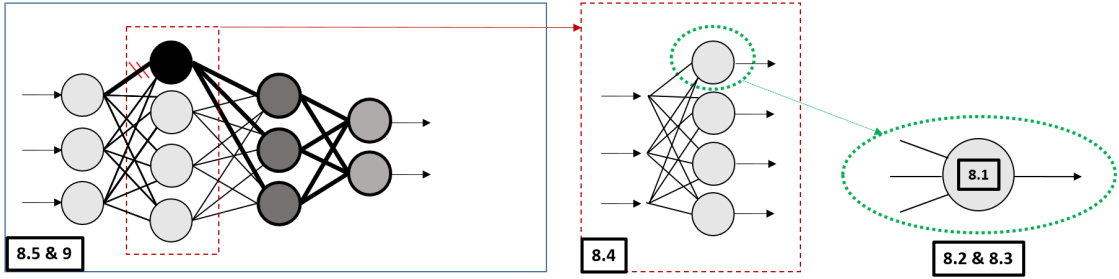}
	\caption{The basic elements of a MLP model studied in the paper with the related Sections (black rectangles). (Left) The snowball effect (Section~\ref{desc_challenges}): an extraction error on a weight (first bold connection) misleads the neuron (black) output, propagates through the model and drastically impacts the recovery of all the other weights of the next neurons (dark gray).}
	\label{paper_experiments_illu}
\end{figure}

In Section~\ref{Background}, we first provide basic deep learning backgrounds that introduce most of our formalism. Related works are presented in Section~\ref{related_works}, followed by an explanation of our positioning and contributions (Section~\ref{contribution}). Details on our experimental setups and comments on our implementations are presented in Section~\ref{expsetup}, before a description of the threat model setting, discussed in Section~\ref{threat_model}. As an introduction to all our experiments, we expose the main challenges related to fidelity-based parameter extraction and describe our overall methodology in Section~\ref{challenges}. Then, in Section~\ref{xtr_method_exp_results}, we detail our extraction method, our experiments and results with a progressive focus on: (1) one multiplication operation, (2) one neuron, (3) sign extraction, (4) several neurons and (5) successive layers. As future works (Section~\ref{future_works}), we discuss the critical case of bias and the scaling up of our approach on state-of-the-art models. Fig.~\ref{paper_experiments_illu} illustrate the structure of the experimental sections with respect to the basic structural elements of a model. Finally, we conclude with possible mitigations.

%% file: src/theorical_background.tex
\subsection{Neural Networks}

\subsubsection{Formalism}
This work is about supervised DNN models. Input-output pairs $(x,y) \in \mathcal{X} \times \mathcal{Y}$ depend on the underlying task. A neural network model $\mathcal{M}_W : \mathcal{X} \rightarrow \mathcal{Y}$, with parameters $W$, predicts an input  $x \in \mathcal{X}$ to an output $\mathcal{M}_W(x) \in \mathcal{Y}$ (e.g., a label for classification task). $W$ are optimized during the training phase in order to minimize a \textit{loss} function that evaluates the quality of a prediction compared to the ground-truth $y$. Note, that a model $\mathcal{M}$, seen as an \textit{abstract algorithm}, is distinguished from its \textit{physical implementations} $M^{*}$, for example embedded models in microcontroller platforms. 
From a pure functional point of view, the embedded models rely on the same abstraction but differ in terms of implementation along with potential optimization processes (e.g., quantization, pruning) to reach hardware requirements (e.g., memory constaints).

\subsubsection{Perceptron}

is the basic functional element of a neural network. The perceptron (also called \textit{neuron} in the paper) first processes a weighted sum of the input with its trainable parameters $w$ (also called \textit{weights}) and $b$ (called \textit{bias}), then non-linearly maps the output thanks to an \textit{activation function} (noted $\sigma$):
\begin{equation}
    a(x)=\sigma\Big(\sum_{j=0}^{n-1}{w_{j}}x_j + b\Big)
\label{eq_perceptron}
\end{equation}
where $x=(x_0,..., x_j, x_{n-1}) \in \mathbb{R}^n$ is the input, $w_{j}$ the weights, $b$ the bias, $\sigma$ is activation function and $a$ the perceptron output. The historical perceptron used the sign function as $\sigma$ but others are available, as detailed hereafter. 

\subsubsection{MultiLayer Perceptrons (MLP)} are \textit{deep} neural networks composed of many perceptrons stacked vertically, called a \textit{layer}, and \textit{multiple layers} stacked horizontally. 
A neuron from layer $l$ gets information form all neurons belonging to the previous layer $l-1$. Therefore, MLP are also called  \textit{feedforward fully-connected} neural networks (i.e, information goes straight from input layer to output one). For a MLP, Equation~\ref{eq_perceptron} can be generalized as: 
\begin{equation}
    a^{l}_j(x)=\sigma\Big(\sum_{i \in (l-1)}{w_{i,j}}a^{l-1}_i + b_j\Big)
\label{eq_mlp}
\end{equation}
\noindent where $w_{i,j}$ is the weight that connects the neuron $j$ of the layer $l$ and the neuron $i$ of the previous layer ($l-1$), $b_j$ is the bias of the neuron $j$ of the layer $l$ and $a^{l-1}_i$ and $a^{l}_j$ are the output of neuron $i$ of layer $(l-1)$ and neuron $j$ of layer $l$.

\subsubsection{Activation functions} inject non-linearity through the layers. Typical functions maps the output of a neuron into a well-defined space like $[0,+\infty]$, $[-1,+1]$ or $[0,1]$.
 
The Rectified Linear Unit function (hereafter, ReLU) is the most popular function because of its simplicity and constant-gradient property.
ReLU is piece-wise linear and defined as $ReLU(x)=max(0,x)$. 
We focus our work on ReLU but other activations are possible: $tanh$ , $sigmoid$ or $softmax$ that is typically used at the end of classification models to normalize output to a probability distribution.

\subsection{IEEE-754 Standard for Floating-Point Arithmetic}
\label{ieee754}
We study single-precision floating-point values on a 32-bit microcontroller. IEEE-754 standard details floating-point representation and arithmetic. Floating value are composed of three parts: Sign, Exponent and Mantissa as in Eq.~\ref{eq_ieee754} for a 32-bit single-precision floating-point value, $a$:

\begin{align}
\label{eq_ieee754}
    a &= (-1)^{b_{31}}\times 2^{(b_{30}...b_{23})_2-127}\times\Big( 1.b_{22}...b_{0} \Big)_2 \\
     &= (-1)^{S_a} \times 2^{E_a - 127} \times \Big(1 + 2^{-23} \times M_a\Big) \nonumber
\end{align}

This allows to represent values from almost $10^{-38}$ to $10^{+38}$ and considers specific case like infinity or \textit{Not a Number} (NaN) values which are not considered here. We emphasize on the \textit{usual case} when the exponent value belongs to $[\![1;254]\!]$. In this case, the final floating-point value $a$ is as in Eq.~\ref{eq_ieee754} where $S_a$, $E_a$ and $M_a$ correspond respectively to the sign, exponent and mantissa values. With this representation, result of the multiplication operation $c = a \times b$ with $b$ another single floating-point value, leads to the sign ($S_c$), exponent ($E_c$) and mantissa ($M_c$) detailed in Eq.~\ref{ieee_754_c}.
Note that these do not necessarily correspond the very final representation of $c$: depending on the value of $M_c$, some realignment can be performed affecting both $M_c$ and $E_c$. However, it appears clearly that $M_a \times M_b$ have less impact on $M_c$ value than $M_a + M_b$.

\begin{align}
\label{ieee_754_c}
S_c &= S_a \oplus S_b\\\nonumber
E_c &= E_a + E_b - 127\\\nonumber
M_c &= M_a + M_b + 2^{-23} \times M_a \times M_b 
\end{align}

%% file: src/related_works.tex
Table~\ref{sum_table} presents works that are~--~to the best of our knowledge~--~references for the topic of model extraction. These works are distinguished through the adversary's objective (recover the architecture or recover the parameters) and the attack surface (API-based attacks or side-channel-based approaches). In this section, we detail works related to our scope. Interested readers may refer to surveys with a wider panorama such as~\cite{mendez2021physical} or~\cite{chabanne2021side}\footnote{More particularly, cache-based attacks that are out of our scope.}.

\input{tab/attack_summary_table}

\subsection{API-based attacks} 
These approaches exploit input/output pairs and information about the target model.     
Carlini \etal~consider the extraction of parameters as a \textit{cryptanalytic} problem~\cite{carlini_cryptanalytic_2020} and demonstrate significant improvements from~\cite{jagielski_high_2020}. The threat model sets an adversary that knows the architecture of the target model but not the internal parameters. The attack is only focused on ReLU-based multi-layer perceptron (MLP) models with one (scalar) output. The basic principle of this attack exploits the fact that the second derivative of ReLU is null everywhere except at a \textit{critical point}, i.e. at the boundary of the negative and the positive input space of ReLU. By forcing exactly one neuron at this critical state thanks to chosen inputs and binary search, absolute values of weight matrix can be reconstructed progressively. Then, the sign is obtained thanks to small variations on the input and by checking activation output. 
Experimental results (state-of-the-art) show a complete extraction of a 100,000 parameters MLP (one hidden layer) with $2^{21.5}$ queries with a worst-case extraction error of $2^{-25}$. Although the attack is an important step forward, limitations rely on its high complexity for deeper models and its strict dependence to ReLU.

\subsection{Timing Analysis}

In \cite{gongye_reverse-engineering_2020}, Gongye~\etal~exploit, on a x86 processor, extra CPU cycles that significantly appear for IEEE-754 multiplication or addition with \textit{subnormal} values. They precisely recover a 4-layer MLP models (weights \textit{and} bias). However, a potential simple countermeasure against this attack is to enable \textit{flush-to-zero} mode which turns subnormal values into zeros.

Maji \etal~also demonstrate a timing-based parameter recovery that mainly rely on ReLU and the multiplication operation~\cite{maji2021leaky} with floating point,
fixed point, and binary models deployed on three platforms without FPU (ATmega-328P, ARM Cortex-M0+, RISC-V RV32IM). Countermeasures encompass adapted floating-point representation and a constant-time ReLU implementation. However, they highlight the fact that even with constant-time implementations, correlation power analysis (CPA) may be efficient and demonstrate a CPA (referencing to~\cite{batina_csi_2019}) on only one multiplication.

\subsection{SCA-based extractions}
\cite{batina_csi_2019} from Batina \etal~, is a milestone work  
that covers the extraction of model's architecture, activation function and parameters with SCA. Two platforms are mentioned, Atmel ATmega328P (opened) and SAM3X8E ARM Cortex-M3 for which floating-point operations are performed without FPU. 

Activation functions are characterized with a timing analysis that enables a clear distinction between \textit{ReLU}, \textit{sigmoid}, \textit{tanh} and \textit{softmax} and relies on the strong assumption that an adversary is capable of measuring precisely execution delay of each activation functions of the targeted model during inference. 

The main contribution, for our work, is related to the parameter extraction method that is mainly demonstrated on the 8-bit ATmega328P. Bias extraction is not taken into account nor mentioned. The method is focused on a low-precision recovery of the IEEE-754 float32 weights. Correlation Electromagnetic Analysis is used to identify the Hamming Weight ($HW$) of multiplication result (\texttt{STD} instructions to the 8-bit registers). The weight values are set in a realistic range $[-N, +N]$ with a precision $p=10^{-2}$ (therefore, $2N/p$ possible values). They extract the three bytes of the mantissa (three 8-bit registers) and the byte including the sign and the exponent\footnote{Due to IEEE-754 encoding, second byte of an encoded value contains the least significant bit of the exponent and the 7 most significant bits of mantissa}.

There is no mention of an adaptation of this technique when dealing with the 32-bit Cortex-M3. Since desynchronization is strong (software multiplication and non-constant time activation function), the EM traces are resynchronized each time according to the target neuron. Note that, because the scope of \cite{batina_csi_2019} also encompass timing-based characterization and structure extraction, the scaling up from one weight to a complete deep model extraction and the related issues are not detailed. 

Finally, model's topology is extracted during the weight extraction procedure: new correlation scores are used to detect layer boundaries, i.e. distinguish if currently targeted neuron belongs to the same layer as previously attacked neurons or to the next one.

Presented methods are confronted to a MLP trained on MNIST dataset and a 8-bit convolutional neural network (CNN) trained on CIFAR-10\footnote{The specific features of CNN compared to MLP that should impact the leakage exploitation are not discussed in~\cite{batina_csi_2019}.}. Original and recovered models have an accuracy difference of 0.01\% and 0.36\% respectively, with an average weight error of 0.0025 for the MLP. Implementations are not available and the compilation level is not mentioned.

%% file: tab/attack_summary_table.tex
\begin{table}[t]
\captionsetup{justification=justified}
 \caption{Related state-of-the-art works. NS: Not Specified, µC: Microcontroller, AR: Architecture Recovery, PR: Parameters Recovery, TA: Timing Attack}
 \label{sum_table}
 \centering
 \resizebox{\linewidth}{!}{%
 \begin{tabular}{lccccr}
  Attack & Target & Technique & AR & PR & Specificity \\

  \toprule  

  Carlini \etal~\cite{carlini_cryptanalytic_2020,jagielski_high_2020}       & N.S. & API & & \checkmark & Target ReLU in MLP \\
  
  Oh \etal~\cite{oh_towards_2018}                       & N.S. & API & \checkmark & & Classifying DNN arch. from querries \\
  
  \midrule

  Gongye \etal~\cite{gongye_reverse-engineering_2020}   & x86 proc. & TA & & \checkmark & Use IEEE754 subnormal values \\
  Maji \etal~\cite{maji2021leaky}                       & µC & TA &  & \checkmark & CNN recovery (1, 8 and 32 bits) \\
  Xiang \etal~\cite{xiang_open_2020}                    & µC & SPA + ML & \checkmark &  & Classifying DNN arch. from traces \\
  Batina \etal~\cite{batina_csi_2019}                   & µC & TA + CPA & \checkmark & \checkmark & Arch. \& low-fidelity param. extraction\\ 
  \textbf{Ours} & \textit{µC} & \textit{CPA} &  & \textit{\checkmark} & \textit{High-fidelity parameters extraction}\\
  \midrule

  Hua \etal~\cite{hua_reverse_2018}                     & FPGA & SPA & \checkmark & \checkmark & Targeting memory-access pattern \\
  Dubey \etal~\cite{dubey_maskednet_2020}               & FPGA & CPA & & \checkmark & Advanced leak-model over BNN \\
  Yu \etal~\cite{yu_deepem_2020}                        & FPGA & CPA + API & \checkmark & & Reconstruct BNN model \\
  
  \midrule
  
  Breier \textit{et al.}~\cite{breier2021sniff}         & N.S. & FIA  & & \checkmark & Extract the last layer\\
  
  \bottomrule
  
 \end{tabular}
}
\end{table}

%% file: src/contribution.tex
\begin{figure}[t]
	\centering
	\includegraphics[width=0.45\textwidth]{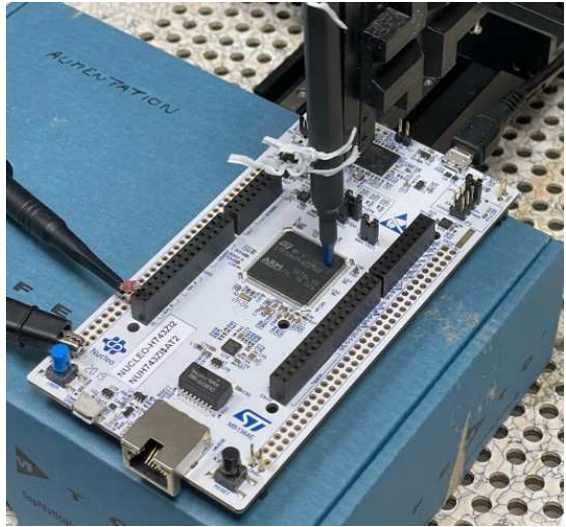}
	\caption{Experimental setup.}
	\label{exp_setup}
\end{figure}

Our scope is a \textit{fidelity-based} extraction of parameters of a MLP model embedded for inference purpose in an AI-suitable 32-bit microcontroller thanks to correlation-based SCA, such as CPA or CEMA. 
Our principal reference is the work from Batina \etal~\cite{batina_csi_2019} (and to certain extend~\cite{carlini_cryptanalytic_2020} as a state-of-the-art fidelity-based extraction approach) and we position our contributions as follow:
\begin{itemize}
    \item Contrary to~\cite{batina_csi_2019} (precision of $10^{-2}$), we set in a fidelity scenario and aim at studying how SCA can precisely extract parameter values. 
    \item Our claim is that the problem of parameter extraction raises several challenges, hardly mentioned in the literature, that we progressively describe. A wrong assumption may reduce this problem to a naive series of attacks targeting independent multiplications (that are actually not independent). 
    \item From the basic operation (multiplication) to an overall model, we propose and discuss methods to extract the complete value of 32-bit floating point weights. Extraction error can reach IEEE-754 encoding error level. 
    \item We do not claim to be able to fully recover all the parameters of a software embedded MLP model: we show that extraction of a secret weight absolute value from multiplication operation is necessary but not sufficient to generalize to the extraction of a complete MLP model. We discuss open issues preventing this generalization such as the extraction of bias values. 
    \item We highlight the choice of our target, based on a ARM-Cortex M7, i.e. a high-end device particularly adapted to deep learning applications (STM32H7). To the best of our knowledge, such a target does not appear in the literature despite its DNN convenient attributes (e.g., FPU, memory capacity). Electromagnetic (EM) acquisitions have been made with an unopened chip which corresponds to a more restrictive attack context compared to literature.
    \item To foster further experiments and help the hardware security community to take on this topic, our traces and implementations are publicly available\footnote{\url{https://gitlab.emse.fr/securityml/SCANN-ex.git}}.
\end{itemize}

%% file: src/experimental_setup.tex
\subsection{Target device and setup}
Our experimental platform is a ARM Cortex-M7 based STM32H7 board. This high-end board provides large memories (2 MBytes of flash memory and 1 MByte of RAM) allowing to embed state-of-the-art models (e.g., 8-bit quantized MobileNet for image classification task). A 25 MHz quartz has been melted as part of the HSE oscillator to have more stable clock. CPU is running at 25 MHz as well, as its clock is directly derived from the melted quartz.
EM emanations coming from the chip are measured with a probe from Langer (EMV-Technik LF-U 2,5 with a frequency range going from 100 kHz to 50 MHz) connected to a 200 MHz amplifier (Fento HVA-200M-40-F) with a 40 dB gain, as shown in Fig.~\ref{exp_setup}. Acquisitions are collected and saved thanks to a Lecroy oscilloscope (4 GHz WaveRunner 640Zi). 

To reduce noise and ease leakage exploitation, all traces acquired experimentally from Cortex M7 are averaged over 50 program executions.
\input{tab/listing_relu_constant_time}
\subsection{Inference program}
Because of the scope, objective and methodology of this work, we need to perfectly master the programs under analysis to properly understand the leakage properties and their potential exploitation. Therefore, instead of attacking black-box off-the-shelf inference libraries,
 we implement our own C programs for every experiments mentioned in this paper 
and compile them with \texttt{O0} gcc optimization level to ensure each multiplication is followed by \texttt{STR} instruction saving result in SRAM. This point is discussed as further works in Section~\ref{mitigation}. 

As in~\cite{maji2021leaky}, some approaches exploit timing inconsistency to recover model information. In this work, we consider implementations protected from such kind of attacks as model inferences are performed in a timing constant manner. We claim that these choices represent more real-world applications, as for the selection of an high-end AI-suitable board:
\begin{itemize}
    \item We use the floating-point unit (FPU) module that performs floating-point calculations in a single cycle rather than passing through C compiler library. When available, usage of such \textit{hardware} module is preferred to its \textit{software} counterpart as it speeds up program execution and relieve CPU. 
    \item ReLU function has been implemented in a timing constant way as in Listing~\ref{relucsttime}. It has been confirmed by checking on thousands of execution that its delay standard deviation is lower than one clock cycle.  
\end{itemize}

%% file: tab/listing_relu_constant_time.tex
\begin{lstlisting}[float=t, language={C} ,frame=single, basicstyle=\ttfamily\scriptsize,  caption=constant-time ReLU implementation ,label=relucsttime]
float layer_neuron_res; // input
int sign,mask,pre_v,post_v;
void *ppre_v,*ppost_v;

sign=(layer_neuron_res>0.0);
mask=0-sign;
ppre_v=(void*)(&(layer_neuron_res));
pre_v=*((int*)ppre_v);
post_v=pre_v & mask;
ppost_v=(void*)(&(post_v));
layer_neuron_res=*((float *)ppost_v); // output
\end{lstlisting}

%% file: src/threat_model.tex
\indent\textbf{Adversary's goal.}
Considered adversary aims at reverse-engineering an embedded MLP model as closely as possible by cloning the targeted model with a fidelity-oriented approach. This objective implies that parameters values resulting from target model training phase have to be estimated. 

\indent\textbf{Adversary's knowledge.}
The attack context corresponds to a gray-box setting since the adversary knows several information about the target system: (1) the model architecture, (2) the used activation function is ReLU, (3) model parameters are stored as single-precision float following the IEEE-754 standard.
With an appropriate expertise in Deep Learning, the attacker may also carry out upstream analyses more particularly on the typical distribution of the weigths (ranges, normalization...) he aims at extracting.

\indent\textbf{Adversary's capacity.}
The adversary is able to acquire side-channel information (in our case, EM emanations with an appropriate probe), leaking from the system embedding the targeted DNN model. The collected traces only results from the usual inference and the attacker does not alter the program execution.  
We assume a classical linear leakage model: the leakage captured is linearly dependent of the $HW$ of the \textit{processed} secret value (e.g., a floating-point multiplication between the secret $w$ and an input coming from the previous layer). Typically, a gaussian noise encompasses the intrinsic and acquisition noise.

The adversary can feed the model with crafted inputs, without any limitation (nor normalization), allowing to control the distribution of the inputs according to the chosen leakage model. However, these chosen inputs belong to the \textit{usual} values according to the IEEE-754 standard. Contrary to API-based attacks, the attacker does not need to access the outputs of the model.

To simplify the scope of this introductory work, we set in a \textit{worst} case scenario according to defender. Attacker is considered able to access a clone device and have enough knowledge and expertise to take benefit of his own implementations to estimate the temporal windows in which he will perform his analysis. We discuss that point in Section~\ref{challenges} and \ref{future_works}.

%% file: figures/traces_nn_depth.tex
\begin{figure}[h]
	\centering
	\includegraphics[width=0.92\textwidth]{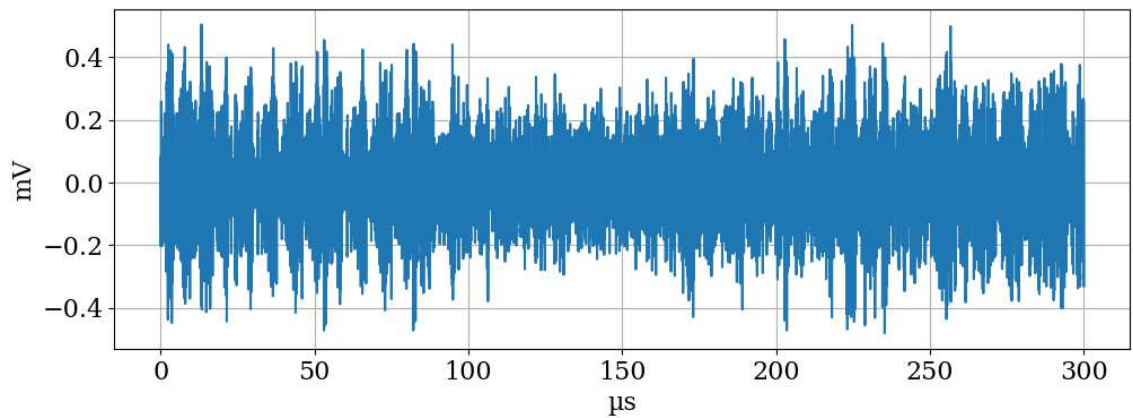}
	\caption{An averaged trace of a 5 layers-deep model composed of 64 positive weights and no bias.} 
	\label{NN_depth_avg_traces}
\end{figure}

%% file: src/introduction_and_problems.tex
\subsection{Critical challenges related to SCA-based parameter extraction}
\label{desc_challenges}

\indent\indent\textbf{Impact of ReLU.} 
The assumption stating that inputs of targeted operation are controlled by the adversary is partially correct: if we focus on a single hidden layer, the inputs are the outputs of the previous one after passing through ReLU. Therefore, the input range is necessarily restricted to non-negative values.  

\indent\textbf{Fully-connected model.} MLP parameters are not shared\footnote{Contrary to Convolutional Neural Network models.}: the activated output of a neuron is connected to \textit{all} neurons of the next layer. Then, an input is involved in as many multiplications as neurons in the considered layer. As such, when performing a CEMA at the layer-level, several hypothesis would stand out from the analysis and would likely be correct as they would correspond to the leakage of each neuron of the layer. However, these hypothesis would not stand out at the same time if neurons outputs are computed sequentially. Therefore, knowledge of the order of the neurons is compulsory to correctly associate CEMA results to neurons. That point is closely related to the threat model we defined in Section~\ref{threat_model} and the profiling ability of the attacker is also discussed in Section~\ref{future_works}.

\indent\textbf{Error propagation problem.}
Because of the feedforward functioning of a MLP, extraction techniques must be designed as well: the correct extraction of parameters related to a layer cannot be achieved without fully recovering the previous parameters. A strong estimation error in the recovery of a weight (and therefore in the estimation of the neuron output) will impact the extraction of remaining neuron weights. The impact of this error will spread to the weight extraction of all neurons belonging to forthcoming layers as illustrated in Fig.~\ref{paper_experiments_illu}.

\indent\textbf{Temporal profiling.} In~\cite{batina_csi_2019}, side-channel patterns could be visually recognized on 8-bit microcontroller on which most of the results have been demonstrated. In our context, SPA is hardly feasible (e.g., see Fig.~\ref{NN_depth_avg_traces}) and the localization of all the relevant parts of the traces is a challenging issue that we consider as out the scope of this work. As mentioned in our threat model (Section~\ref{threat_model}), we set in a \textit{worst case} scenario where the attacker is able to perform a \textit{temporal profiling} on a clone device to have an estimation of the parts of the traces to target since he has several secrets to recover spread all over the traces. This estimation can be more or less precise to enable attacks at neuron or layer-level.       

\indent\textbf{Bias values.}
Knowledge of the bias value is compulsory to compute the entire neuron output. This parameter is not involved in multiplications with the inputs but added to the accumulated sum between neuron inputs and its weights. Thus, leakage of bias and weight must be exploited differently. Bias extraction is treated in the API-based attack~\cite{carlini_cryptanalytic_2020} and the timing attack from~\cite{maji2021leaky} but not mentioned in~\cite{batina_csi_2019}. In this work, we clearly states that we keep the extraction of bias values as a future work but discuss this challenging point in Section~\ref{what_about_bias}. 

\subsection{Our methodology}

Our methodology starts with analyzing the most basic operation of a model~--~i.e. a multiplication~--~then, to widen our scope, to a neuron, one layer, then several layers as illustrated in Fig.\ref{paper_experiments_illu}. Corresponding steps are evaluated with both simulated and real traces.
Dealing with an entire model means to recover parameters layer by layer, following the (feedforward) network flow: extractions of a layer $l$ being used to infer the inputs of the layer $l+1$.

Since our main objective is a practical fidelity-based extraction, we aims at crafting an efficient extraction method, faster than a \textit{brute-force} CEMA on $2^{32}$ hypothesis, that enables a progressive precision. With this introduction work, in addition to expose challenges that suffer analysis in the literature, we assess the precision degree SCA can reach.        

%% file: src/Experiences_Results/extraction_process.tex
\subsection{Targeting multiplication operation}
\label{targeting_multiplication}

We first focus on the multiplication $c=w \times x$, between two IEEE-754 single-precsion floating-point operands: a secret weight ($w$) and a known input ($x$). We remind that we use hardware operations thanks to the FPU.

\subsubsection{Our approach}
is composed of multiple CEMA to extract the absolute value $|w|$. Importantly, the sign bit is not considered yet and is extracted later on (Section~\ref{targeting_sign}). 
With \textit{fidelity-oriented} extraction as objective, our method has no fixed accuracy objective and avoids exhaustive analysis over $2^{32}$ hypothesis. It allows to see how accurate SCA-based extraction can perform.
It relies on two successive steps. First one tends to recover as much information as possible in a single attack by targeting most significant bits from a variable encoded with IEEE-754 standard. The second step is made to correct possible approximations from the previous one and enhance extraction accuracy by refining the granularity of tested hypothesis.
In this step, no focus is made on specific bits, entire variable with all 32-bit varying are considered. Fig.~\ref{xtr_process_flowchart} describes these two steps. The attack relies on different parameters:
\input{figures/xtr_process_illu}

\begin{itemize}
    \item $d_0$: size of the initial interval that is centered on the value $C$. Thus, the tested hypothesis belong to $[C-d_0/2,C+d_0/2]$. 
    \item $\lambda_1>\lambda_2$: two shrinking factors that narrow the interval of analysis ($\lambda_1$ for the first iteration of step 2, $\lambda_2$ for successive step 2 iterations).
    \item $m$: number of times the step 2 is repeated.
    \item $N$: number of kept hypothesis (after CEMA) at each extraction step.
\end{itemize}

\textbf{Step 1} of the extraction process is as follows: 
\begin{enumerate}
    \item First, we generate an exhaustive set of hypothesis with all possible $2^{16}$ combinations of the 8 bits of exponent and the 8 most significant bits of mantissa (remaining bits are set to 0) and filter out unlikely values (in a DNN context) by keeping hypothesis in $[C-d_0/2,C+d_0/2]$. Kept hypothesis are not linearly distributed in this interval. 
    \item We compute the HW of the targeted intermediate values: here, the result of the products between inputs and weight hypothesis.
    \item We perform a CEMA between EM traces and our HW hypothesis and keep the $N$ best ones according to the absolute values of correlation scores.
\end{enumerate}

\textbf{Step 2} is processed in an iterative way and depends on the previous extraction that could be the output of Step 1 or from the previous Step 2 iteration:
\begin{enumerate}
    \item For each best hypothesis $\hat{w}_i$ kept from the previous step ($i \in \llbracket 0;N-1 \rrbracket$), a set of assumptions is linearly sampled around $\hat{w}_i$ with an narrower interval of size $d_1=d_0/\lambda_1$ (if the previous step was step 1) or $d_{i+1}=d_i/\lambda_2$ (otherwise).
    \item As in step 1, we compute the HW of the intermediate values and perform a CEMA to select the $N$ best hypothesis among the $N$ considered hypothesis sets (so that we always keep $N$ hypothesis at each iteration) according to absolute value of the correlation scores.
\end{enumerate}

Fig.~\ref{xtr_process_illu} shows the two steps of this extraction process ($w=0.793281$, $d_0 = 5$ and $C = 2.5$) for 3,000 real traces on a STM32H7, obtained from the $\texttt{PRGM}_2$ experiment described hereafter. The second step is iterated three times and we progressively reach a high correlation score. 

\subsubsection{Validation on simulated traces}
We first confirm our approach on simulated traces by computing the success rate of our extraction with respect to several extraction error ($\epsilon_{rr}$) thresholds. We randomly generate 5,000 positive secret values $w$ and for each of them, we craft 1,000 3-dimensional traces using random inputs $x$. At the middle sample, the trace value is the Hamming Weight of the multiplication: $HW(x \times w)$. A random uniform variable is used for the other samples. An additional gaussian noise ($\mu=0$, $\sigma$) is applied on the entire trace. We set $N=5$, $d_0=5$, $C=2.5$, $m=3$, $\lambda_1=100$ and $\lambda_2=50$. Results according to the noise level are presented in Tab.~\ref{simu_xtr_process_success_rate_table}. We reach a significant success rate over 90\% for the extraction process until a recovering error of $10^{-6}$.

\input{tab/tab_simu_1_multiplication}

\subsubsection{Experiments on Cortex-M7}
\input{figures/traces_2diffmul}

This extraction method is also confronted to real traces obtained from our target board. For these experiments, the secret values are positive and hard-written in the code and input values are sent from a python script through UART interface, 150,000 traces have been acquired for each of them, then averaged to 3,000 traces. Two programs have been implemented: $\texttt{PRGM}_1$ performs a single multiplication and $\texttt{PRGM}_2$ performs two multiplications with distinct secret values and inputs (corresponding EM activity is depicted in Fig.~\ref{early_trace_2_diff_mul}).
Both being compiled with \texttt{O0} gcc optimization level, this implies that each multiplication is followed by a \texttt{STR} instruction saving the multiplication result in SRAM as in Listing~\ref{assembler_multiplication}. One source of leakages exploited to recover secret values is these store instructions. Inference EM activity to be analyzed is framed by a trigger added by hand at assembler level. 
For this experiment, our extraction method allows to recover the secret values with high extraction level as presented in Tab.~\ref{early_traces_extractions}. 

\begin{table}[h]
    \begin{minipage}{0.48\textwidth}
    \lstset{linewidth = 0.9\linewidth, breaklines=true}
    \input{tab/listing_FPU_multiplication}

    \end{minipage}
    \qquad
    \begin{minipage}{0.48\textwidth}
    \caption{Extraction results targeting multiplications.}
    \label{early_traces_extractions}
    \centering
    \begin{tabular}{ cccc }
    \toprule
    Program & Correct Value &  $\epsilon_{rr}$ \\ 
    \midrule
    $\texttt{PRGM}_1$      & 0.793281 & 4.09e-08  \\ 
    \midrule
    \multirow{2}{*}{$\texttt{PRGM}_2$} & 0.793281 & 1.87e-08  \\ 
                                       & 0.33147  & 1.27e-08  \\ 
    \bottomrule
    \end{tabular}

    \end{minipage}
\end{table}

%% file: figures/xtr_process_illu.tex
\begin{figure}[t]
\centering
\begin{subfigure}[t]{0.45\textwidth}
    \includegraphics[width=0.95\textwidth]{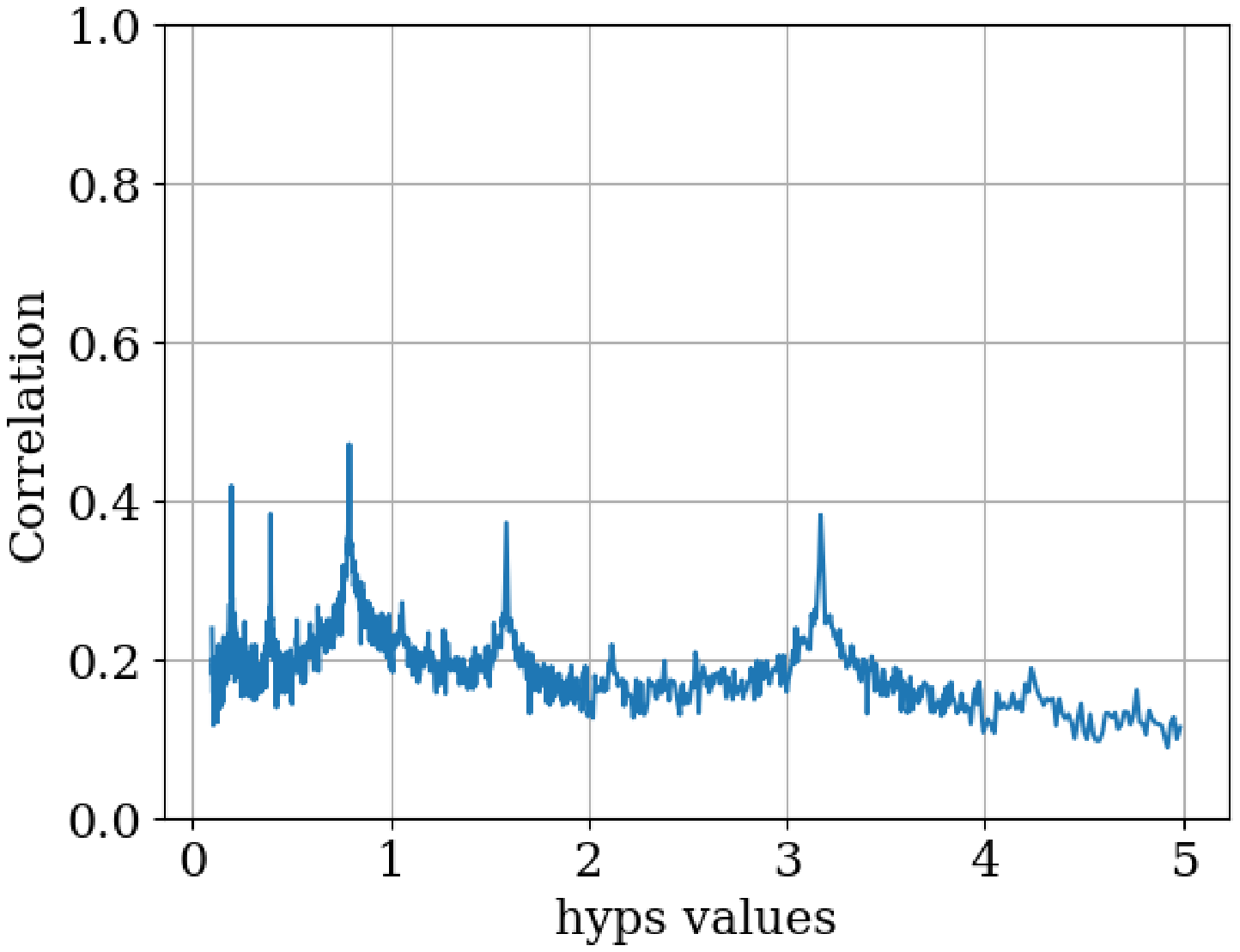}
    \caption{Step 1}
\end{subfigure}\hspace{\fill} 
\begin{subfigure}[t]{0.45\textwidth}
    \includegraphics[width=0.95\linewidth]{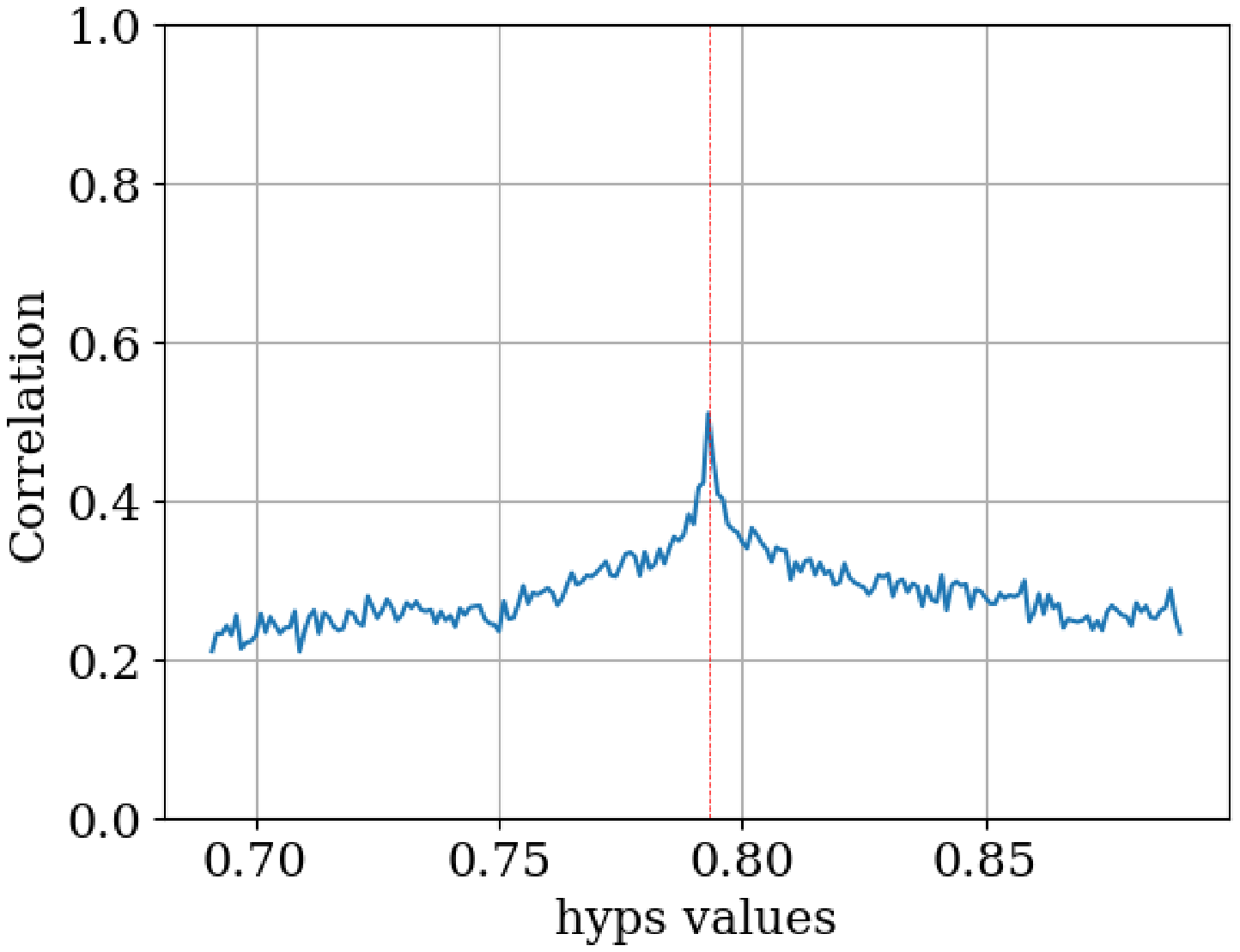}
    \caption{Step 2 iteration 1}
\end{subfigure}

\begin{subfigure}[t]{0.45\textwidth}
    \includegraphics[width=0.95\linewidth]{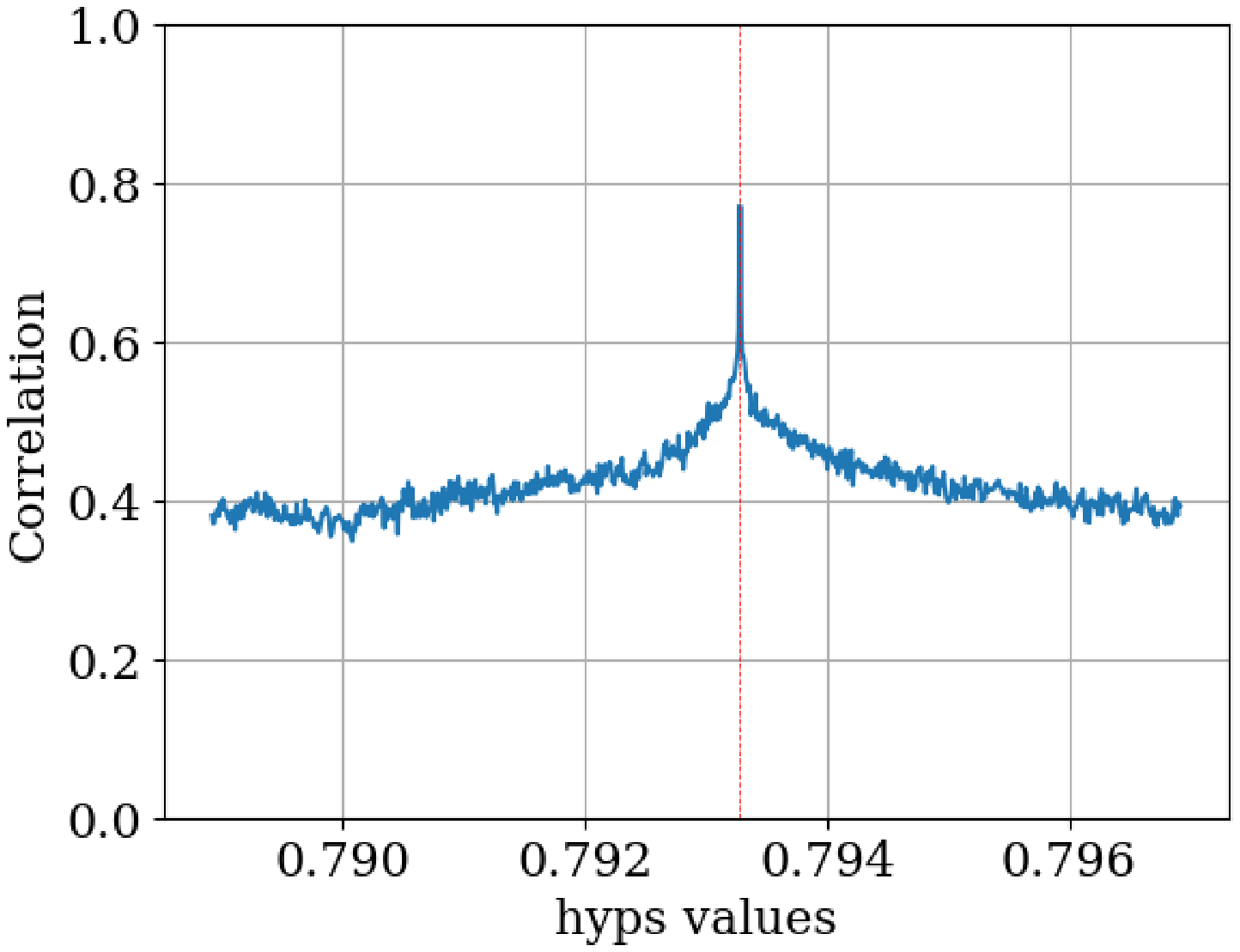}
    \caption{Step 2 iteration 2}
\end{subfigure}\hspace{\fill} 
\begin{subfigure}[t]{0.45\textwidth}
    \includegraphics[width=0.95\linewidth]{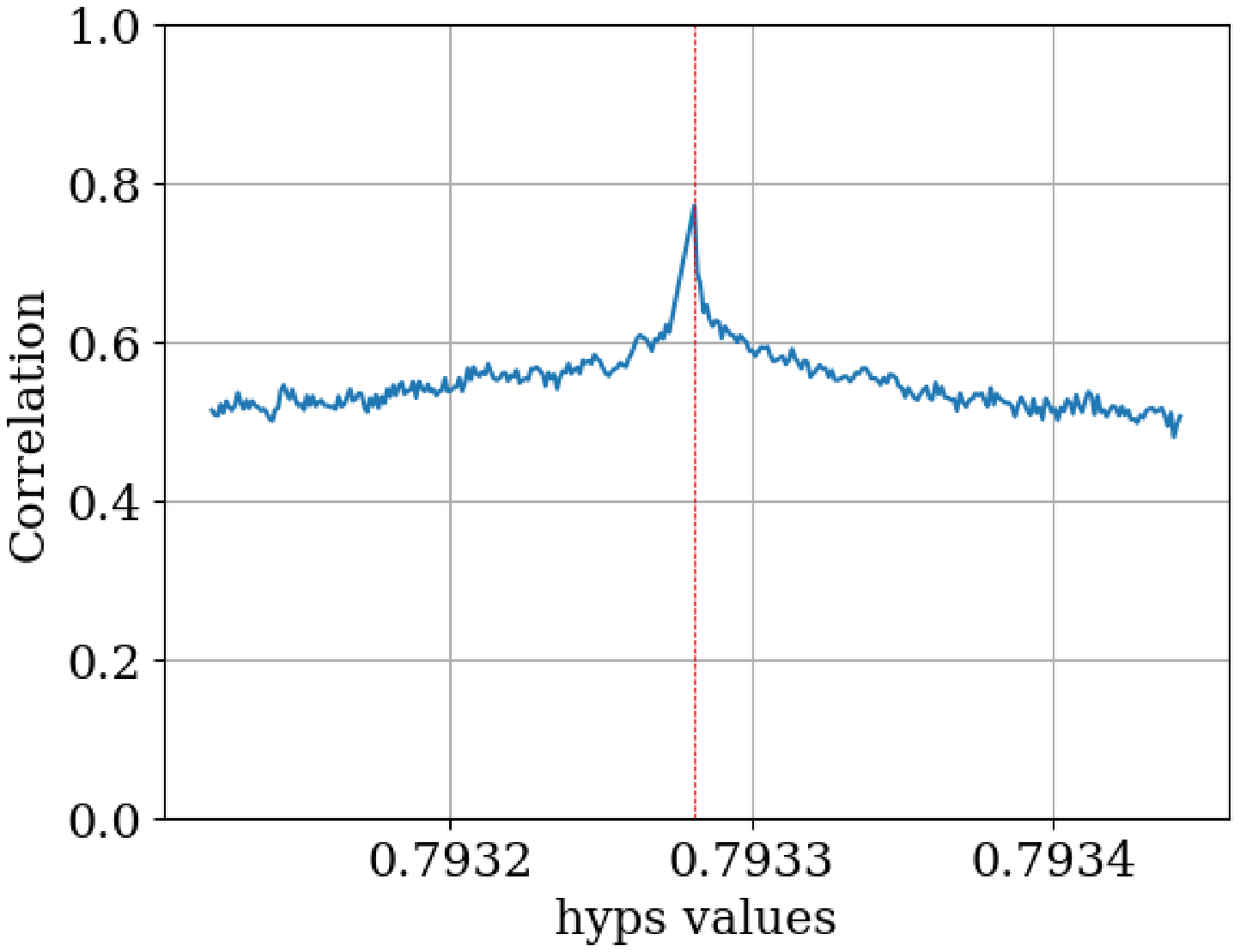}
    \caption{Step 2 iteration 3}
\end{subfigure}

\caption{Our extraction method applied to traces from $\texttt{PRGM}_2$. Step 1 (a) is focused on 8 bits of exponent and 8 bits of mantissa, then Step 2 (b-d) is repeated 3 times.}
\label{xtr_process_illu}
\end{figure}

%% file: tab/tab_simu_1_multiplication.tex
\begin{table}[b]
\caption{Attacking a single multiplication on simulated traces. Success-rate (SR) of the extraction ordered by estimation error ($\epsilon_{rr}$) thresholds according to the noise level ($\sigma$).}
\centering
\label{simu_xtr_process_success_rate_table}

 \addtolength\tabcolsep{5pt}
    \begin{tabular}{clcccccccc}
        \toprule
    \multicolumn{2}{c}{$\epsilon_{rr}\leq$} & $10^{-1}$ & $10^{-2}$ & $10^{-3}$ & $10^{-4}$ & $10^{-5}$ & $10^{-6}$ & $10^{-7}$ & $10^{-8}$\\
        \midrule 

        \multirow{4}{*}{SR} & $\sigma^2=0.5$ & 100 & 100 & 98.4 & 98.3 & 96.4 & 94.2 & 81.8 & 77.1 \\

        & $\sigma^2=1$ & 100 & 99.9 & 98.8 & 98.6 & 96.9 & 94.8 & 81.2 & 75.4 \\

        & $\sigma^2=25$ & 99.9 & 99.2 & 97.0 & 96.8 & 94.9 & 91.6 & 78.0 & 73.2 \\

        & $\sigma^2=10^2$ & 99.9 & 98.2 & 94.3 & 93.0 & 89.8 & 86.5 & 69.6 & 62.4 \\
        \bottomrule
    \end{tabular}
\end{table}

%% file: figures/traces_2diffmul.tex
\begin{figure}[t]
    \centering
    \includegraphics[width=0.8\linewidth]{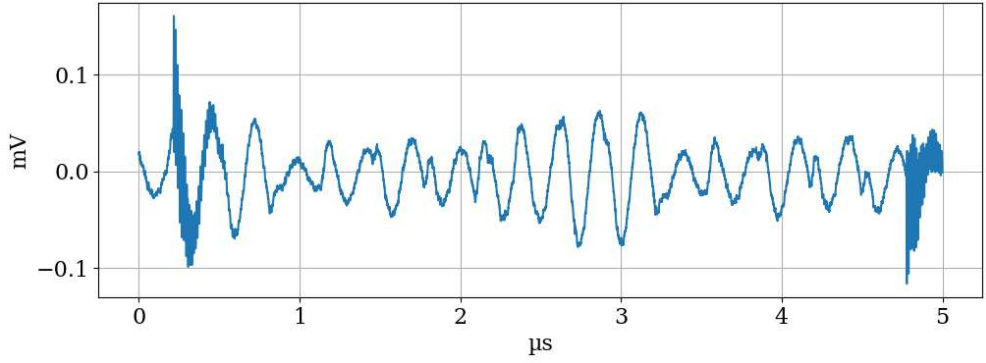}
    \caption{Averaged trace acquired when observing $\texttt{PRGM}_2$ execution.}
    \label{early_trace_2_diff_mul}
\end{figure}

%% file: tab/listing_FPU_multiplication.tex
\begin{lstlisting}[language={[x86masm]Assembler} ,frame=single, basicstyle=\ttfamily\scriptsize, caption=Assembler code of multiplication execution using FPU ,label=assembler_multiplication]
; load input
vldr     s14, [r7, #12]
ldr      r3, [pc, #76]
; load weight
vldr     s15, [r3, #4]
; perform product
vmul.f32 s15, s14, s15
; store result
vstr     s15, [r7, #24]
\end{lstlisting}

%% file: src/Experiences_Results/targeting_single_neuron.tex
\subsection{Extracting parameters of a perceptron}

\subsubsection{Neuron computation implementation} 
After extracting secret value from an isolated multiplication and studying success-rate of such attack, we scale up to a single neuron computation as described in Eq.~\ref{eq_perceptron}. The most important difference is that the output of a neuron is the result of an accumulation of several multiplications. 
This accumulation is processed through two successive FPU instructions (\texttt{fmul} then \texttt{fadd}, as it is the case in our experiments with Listing~\ref{assembler_multiplication_accumulation}) or a dedicated multiplication-addition instruction.    

That leads to two new challenges in the extraction of the secret values compared to our first experiments on isolated multiplications: (1) hypothesis have to be made on accumulated values, (2) the attacker needs to know the order in which multiplication are computed (i.e., how the accumulation evolves).

\input{tab/listing_FPU_multiplication_accumulation}

\subsubsection{Extraction of neuron weights} 
We assume that the attacker knows the computation order. The first weight $w_0$ can be extracted as done before by exploiting the direct result of $w_0\times x_0$. Then, the second weight $w_1$ can also be extracted with the same approach by targeting $w_0\times x_0 + w_1\times x_1$ because $w_0$ was recovered before.
This process can be applied again for each weight value as long as all previous ones have been correctly extracted. Actually, that point is a critical one since the extraction quality of currently targeted weight  strongly depends on the extraction accuracy of previously extracted weights.

\subsubsection{Experiments on Cortex-M7}
We apply that method on 2,000 averaged traces that capture the inference of a 4-input neuron. 
As presented in Table~\ref{neuron_traces_extractions}, we reach a very precise extraction of the four weights.
\input{tab/4input_neuron_traces_extraction}

\subsection{Targeting the sign}
\label{targeting_sign}
\subsubsection{Problem statement} 
As seen before, for a ReLU-MLP model, a neuron belonging to an hidden layer is fed with non-negative input values. An obvious but important observation is that, for $w \times x = c$, if the secret value $w$ is multiplied with positive value $x \geq 0$ then $sign(c)=sign(w)$. Therefore, in such a context, CEMA is not able to distinguish sign by leveraging the input-weight product. 

\subsubsection{Extracting the sign at the neuron-level} 
A way to overcome this issue is to set the sign extraction problem at the neuron-level, i.e. to build hypothesis on sign changes throughout the overall multiplication-accumulation process. 

Let consider the accumulation of two successive multiplications: $acc = w_0 \times x_0 + w_1 \times x_1$ with inputs $x_0, x_1 \geq 0$. $acc$ variations would change whether $sign(w_0) \neq sign(w_1)$ or $sign(w_0) = sign(w_1)$. Based on that, by focusing on variation of $|acc|$ value, it is possible to find out if $w_0$ and $w_1$ have an opposed sign or not.
Thus, weight sign estimation can be done progressively, by checking if the sign associated to the weight currently extracted is similar or opposed to the sign of the previous weight. However, since the sign extraction is processed \textit{relatively} to the sign of $w_0$, an additional verification has to be done to confirm which of the current extracted signs or the opposite is correct. This can be done thanks to ReLU output that matches with only one hypothesis.

\input{tab/simu_signed_weights}

\subsubsection{Validation on simulated traces} 
 As in Section~\ref{targeting_multiplication}, we craft simulated 50-dimensional traces for a $m$-input neuron with $m$ randomly picked in $\llbracket 2;8 \rrbracket$. We generate 5,000 neurons with $m$ signed weights, no bias and fed by 3,000 positive inputs sets (i.e., 25M of traces). The generation process is similar to the previous experiment apart from the leakage placement which depends on $m$. Thus, leakages correspond to $m$ product accumulations and one for the $ReLU$ output. We uniformly place these $m+1$ leakage samples in the traces with random uniform values for the other samples. To characterize the principle of the method, we set in a low-noise simulation ($\sigma^2=0.5$). We reach the following results:
\begin{itemize}
    \item 78.8\% of neurons have been extracted with all signs correctly assigned.
    \item For 91.6\% of the weights, the sign is correctly assigned. Table~\ref{simu_xtr_signed_neuron_success_rate_table} details the extraction success rates for these weights (consistent with Table~\ref{simu_xtr_process_success_rate_table}).  
\end{itemize}
 
\subsubsection{Experiments on Cortex-M7} We use 5,000 averaged traces capturing the inference of one neuron with signed weights. With Table~\ref{signed_neuron_traces_extractions}, we observe that the sign inversion and the relative value have been correctly affected. 
In addition to our previous experiences, our approach performs well at the neuron-level. We progressively scale-up in the next section at a layer-level.  

\input{tab/real_signed_weights}

%% file: tab/listing_FPU_multiplication_accumulation.tex
\begin{lstlisting}[float=h!, language={[x86masm]Assembler} ,frame=single, basicstyle=\ttfamily\scriptsize, caption=Assembler code of multiplication and accumulation  using FPU ,label=assembler_multiplication_accumulation]
 ;layer1_neuron_res[i] += input[j] * weight_layer1[i][j];
 [...]                ; r3 = accumulator address (SRAM)
 vldr      s14, [r3]  ; Load accumulator
 [...]                ; r3 = input address (SRAM)
 vldr      s13, [r3]  ; Load input
 [...]                ; r3 = weight address (FLASH)
 vldr      s15, [r3]  ; Load weight
 vmul.f32  s15, s13, s15  ; Multiplication (FPU)
 vadd.f32  s15, s14, s15  ; Addition (FPU)
 [...]                ; r3 = accumulator address (SRAM)
 vstr      s15, [r3]  ; store result
\end{lstlisting}

%% file: tab/4input_neuron_traces_extraction.tex
\begin{table}[t]
\caption{Extraction error ($\epsilon_{rr}$) from a 4-input neuron on a Cortex-M7 target (3,000 averaged traces)}
\label{neuron_traces_extractions}
\centering
\addtolength\tabcolsep{5pt}
\begin{tabular}{ccc} 
  \toprule

  Target weight & Correct Value & $\epsilon_{rr}$ \\
  \midrule
  $w_0$ &   0.366193473339   & 5.96e-08  \\ [0.5ex]
  $w_1$ &   0.90820813179    & 3.58e-07  \\ [0.5ex]
  $w_2$ &   0.522847533226   & 5.96e-08  \\ [0.5ex]
  $w_3$ &   0.00123456       & 4.21e-08  \\ [0.5ex]
  \bottomrule
\end{tabular}
\end{table}

%% file: tab/simu_signed_weights.tex
\begin{table}[h!]
    \centering
    \caption{Attacking a neuron with simulated traces. Success-rate (SR) of the extraction of weights (with correctly recovered sign) ordered by estimation error ($\epsilon_{rr}$) thresholds.    }
    \label{simu_xtr_signed_neuron_success_rate_table}
    \addtolength\tabcolsep{5pt}
    \begin{tabular}{ccccccccc}
        \toprule

        $\epsilon_{rr}\leq$ & $10^{-1}$ & $10^{-2}$ & $10^{-3}$ & $10^{-4}$ & $10^{-5}$ & $10^{-6}$ & $10^{-7}$ & $10^{-8}$  \\
        \midrule 
        
        SR & 99.9 & 99.1 & 96.2 & 92.8 & 86.5 & 79.3 & 65.4 & 61.0  \\
        \bottomrule
    \end{tabular}
\end{table}

%% file: tab/real_signed_weights.tex
\begin{table}[h]
\caption{Attacking a neuron on Cortex-M7. Extraction error ($\epsilon_{rr}$) for 4 weights.}
\label{signed_neuron_traces_extractions}
\centering
\addtolength\tabcolsep{5pt}
\begin{tabular}{cccc} 
  \toprule
  Target weight & Correct Value & $\epsilon_{rr}$ & Sign match \\
  \midrule
  $w_0$ &   -0.813444  & 1.38e-07 & \checkmark \\ 
  $w_1$ &   0.0671324  & 3.88e-08 & \checkmark \\ 
  $w_2$ &   0.107843   & 2.34e-07 & \checkmark \\ 
  $w_3$ &   0.604393   & 6.50e-08 & \checkmark \\ 
  \bottomrule
\end{tabular}
\end{table}

%% file: src/Experiences_Results/targeting_few_neurons.tex
\subsection{Targeting one layer}
\label{one_layer_section}

Previous structure has inputs involved in only one multiplication with weights. However, neural network interconnections between layers implies that an input value is passed to each neuron of the layer 
and thus is involved in several multiplications with different weights. If neurons are computed sequentially, this means that CEMA would likely bring out several hypothesis that would match weights of different neurons that leak at different moments.

In this context, to associate the extracted values to a specific neuron, we assume that neurons computation is made sequentially from top to bottom of the layer. 
To ensure an already extracted value is not associated again to another neuron, leaking time sample of tested hypothesis are filtered. Consider only leaking sample greater than the one from last extracted value prevents this.

\subsubsection{Experiments on Cortex-M7}
Two experiments have been made:

\begin{enumerate}
\item \textit{2-neuron layer with 3 inputs each}: the 6 weights are positives and 3,000 traces have been captured by feeding the layer with random positive inputs (as for an hidden layer). The six weights have been recovered with an averaged estimation error $\epsilon_{rr} = 2.68e^{-6}$ (worst: $5.18e^{-6}$, best: $4.48e^{-8}$). 
\item \textit{5-neuron layer with 4 inputs each}: the 20 weights encompass positive and negative values and 5,000 traces have been acquired by feeding the layer with random positive or negative values (as for an input layer). We reach a similar extraction error $\epsilon_{rr}=1.03e^{-6}$ (worst: $3.10e^{-6}$, best: $1.55e^{-7}$). All weight signs have been correctly guessed.
\end{enumerate}

%% file: src/Experiences_Results/targeting_few_layers.tex
\subsection{Targeting few layers}

DNN are characterized by layers stacked horizontally. Proposed method is able to extract weights from one layer and is supposed to be applied to each of them one after the other, by progressively reconstructing intermediate layer outputs.

\subsubsection{Experiments on Cortex-M7} 
To verify this principle, we craft a MLP with 5 hidden layers with respectively 5, 4, 3, 2 and 3 neurons. The 64 corresponding weights are positive and the model is fed with 4-dimensional positive inputs. Every weights have been recovered with an estimation error lower than $10^{-6}$ ($SR=95.31\%$ for $\epsilon_{rr}<10^{-7}$, best $\epsilon_{rr}=7.63e^{-10}$, worst $\epsilon_{rr}=6.67e^{-6}$).
Note that sign is not considered in this experiments because the tested model has been crafted and is not functional (i.e., not the result of a training process). Such \textit{ none functional} models are likely to present too many \textit{dead neurons} and even \textit{dead layers} because of the accumulated ReLU effect. The scaling-up to a fully functional state-of-the-art model with signed weights \textit{and} biases is planned for future works and discussed in the next section.

%% file: src/future_works.tex
\input{src/Experiences_Results/targeting_bias}

\subsection{Targeting state-of-the-art functional models}

Further experiments will encompass compressed embedded models thanks to deployment libraries (e.g., TFLite, NNoM) with a focus on Convolutional Neural Network (CNN) models. Indeed, for memory constrains, deep embedded models in 32-bit microcontrollers are usually stored with parameters quantized in 8-bit integers with training-aware or post-training quantization methods. For the most straightforward quantization and embedding approaches, this quantization should simplify the extraction process with only $2^{8}$ hypothesis for each weight and bias as well as an additional extraction of a scaling factor that enables the mapping from 8-bit to 32-bit values. Regarding CNN, these models also rely on multiplication-accumulation operations (and the same activation principle), but the fact that parameters are shared across the inputs should interestingly impact the way leakages could be exploited for a practical extraction.

%% file: src/Experiences_Results/targeting_bias.tex
\subsection{What about neuron bias?}
\label{what_about_bias}
So far, biases have not been considered even though these values may significantly impact neuron outputs and also the way a neuron is implemented: the weighted sum between weights and inputs could be initialized to $0$ or directly to the bias. In the latter case, our extraction method cannot be directly applied and needs an initial and challenging bias extraction based on IEEE-754 addition. 

To better explain this challenge, lets consider that the accumulation is well initialised to 0 (i.e, bias is added \textit{after} the weighted sum). Using simulated traces, we perform our extraction method to recover the weights \textit{and} the bias by focusing on the final accumulation $\sum_j (w_{i,j}\times x_j) + b$. 

5,000 neurons with $m$ secret weights $w_{0..m}$ ($m$ is randomly picked in $\llbracket 2;8 \rrbracket$) and one secret bias $b$ have been generated. 5,000 simulated traces have been crafted for each neuron with random positive inputs. Success rates (SR) are presented in Table~\ref{simu_xtr_biased_neuron_success_rate_table}. These SR only concern weight and bias for which sign has correctly been recovered. This corresponds to $93.25\%$ over 24956 attacked weights and $92.14\%$ over 5000 bias. While SR related to weight extraction remains consistent with previous simulations, SR corresponding to bias extraction significantly drops (e.g., $SR=35.8$ for $10^{-3}$).

A possible explanation relies on the IEEE-754 addition that requires a strict exponent realignment contrary to multiplication. If $a \gg b$ then $a + b = a$ because $b$ value \textit{disappears} in front of $a$. In our context, as inputs $x$ (controlled by the attacker that aims at covering as much as possible the float32 range) are defined randomly, then multiplied by weights, it is likely that $\sum_{j=0}^{n}{w_{j}}x_j \gg b$. Thus, secret information related to bias could be hardly recovered by exploiting our EM traces. Therefore, we need to develop a different strategy (including a coherent selection of the inputs) to exploit potential IEEE-754 addition leakages.

\input{tab/simu_xtr_biased_neuron_success_rate_table}

%% file: tab/simu_xtr_biased_neuron_success_rate_table.tex
\begin{table}[h]
    \centering
    \caption{Neurons extraction (signed weight \& bias) success-rate}
    \label{simu_xtr_biased_neuron_success_rate_table}
     \addtolength\tabcolsep{5pt}
    \begin{tabular}{ccccccccc}
        \toprule
        $\epsilon_{rr}\leq$ & $10^{-1}$ & $10^{-2}$ & $10^{-3}$ & $10^{-4}$ & $10^{-5}$ & $10^{-6}$ & $10^{-7}$ & $10^{-8}$ \\
        \midrule 
        SR weight & 99.9 & 99.5 & 96.8 & 93.7 & 87.2 & 79.8 & 64.9 & 61.3 \\
        SR bias   & 81.7 & 56.3 & 35.8 & 19.7 & 7.44 & 2.6  & 0.7  & 0.2 \\
        \bottomrule
    \end{tabular}
\end{table}

%% file: src/conclusion.tex
Side-channel analysis is a well-known, powerful, mean to extract information from an embedded system. However,
with this work, we clearly question the practicability of a complete parameters extraction with SCA when facing state-of-the-art models \textit{and} real-world platforms. By demonstrating promising results on a high-end 32-bit microcontroller on a high fidelity-based extraction scenario, we do not claim this challenge as \textit{impracticable} but we aim at inciting further (open) works focused on the exposed challenges as well as bridging different approaches with combined API and SCA-based methods. 

An additional outcome from our experiments concerns defenses. Classical \textit{hiding} countermeasures, already demonstrated in other context (e.g., cryptographic modules), should be relevant (as also mentioned in~\cite{batina_csi_2019}). More precisely, randomizing multiplication and/or accumulation order (including the bias) should significantly impact an adversary. An efficient complementary defense could be to randomly add fake or neutral operations at a neuron or layer-level. We keep as future works, the proper evaluation of such state-of-the-art protections in a model extraction context.

%% file: src/acknowledgements.tex
This work is supported by (CEA-Leti) the European project InSecTT (ECSEL JU 876038)\footnote{\url{www.insectt.eu}, InSecTT: ECSEL Joint Undertaking (JU) under grant agreement No 876038. The JU receives support from the European Union’s Horizon 2020 research and innovation program and Austria, Sweden, Spain, Italy, France, Portugal, Ireland, Finland, Slovenia, Poland, Netherlands, Turkey. The document reflects only the author’s view and the Commission is not responsible for any use that may be made of the information it contains.} 
and by the French ANR in the framework of the \textit{Investissements d’avenir} program (ANR-10-AIRT-05, irtnanoelec);  and (Mines Saint-Etienne) by the French program ANR PICTURE (AAPG2020). This work benefited from the French Jean Zay supercomputer with the AI dynamic access program.